\documentclass[prd,letterpaper,onecolumn,amsmath,amsfonts,amssymb,nofootinbib]{revtex4}
\pdfoutput=1
\usepackage {amsmath} 
\usepackage[dvips]{graphicx}        
\usepackage{feynmp}
\usepackage{amssymb}      
\usepackage{latexsym}
\usepackage{tensor}

\newcommand{\fr}{\frac}
\newcommand{\bea}{\begin{eqnarray}}
\newcommand{\eea}{\end{eqnarray}}

\newcommand{\lt}{\left}
\newcommand{\rt}{\right}

\newcommand{\pt}{\partial}

\newcommand{\be}{\begin{equation}}
\newcommand{\ee}{\end{equation}}
\newcommand{\bear}{\begin{eqnarray}} 
\newcommand{\eear}{\end{eqnarray}}

\newcommand{\lapproxeq}{\lower .7ex\hbox{$\;\stackrel{\textstyle
<}{\sim}\;$}}
\newcommand{\gapproxeq}{\lower .7ex\hbox{$\;\stackrel{\textstyle
>}{\sim}\;$}}
\newcommand{\stackdown}[2]{\lower 1.4ex\hbox{$\;\stackrel{\textstyle{#1}}
{\scriptstyle{#2}}\;$}}

\newcommand{\ba}{\begin{eqnarray}}
\newcommand{\ea}{\end{eqnarray}}
\makeatletter
\def\slash{\@ifnextchar[{\fmsl@sh}{\fmsl@sh[0mu]}}
\def\fmsl@sh[#1]#2{%
  \mathchoice
    {\@fmsl@sh\displaystyle{#1}{#2}}%
    {\@fmsl@sh\textstyle{#1}{#2}}%
    {\@fmsl@sh\scriptstyle{#1}{#2}}%
    {\@fmsl@sh\scriptscriptstyle{#1}{#2}}}
\def\@fmsl@sh#1#2#3{\m@th\ooalign{$\hfil#1\mkern#2/\hfil$\crcr$#1#3$}}
\makeatother
\usepackage{color}

\definecolor{orange}{rgb}{0.9,0.2,0}
\definecolor{brown}{rgb}{0.7,0.3,0.2}
\definecolor{fuxia}{rgb}{1,0,1}
\definecolor{skyblue}{rgb}{0,0.1,0.9}
\definecolor{violetred}{rgb}{0.8,0.13,0.56}
\definecolor{deeppink}{rgb}{1.00,0.08,0.5}
\definecolor{pink}{rgb}{1.00,0.75,0.80}
\definecolor{orchid}{rgb}{0.85,0.44,0.84}
\definecolor{lightpink}{rgb}{1.00,0.71,0.76}
\definecolor{bluish}{rgb}{0,0.6,0.8}  

\begin{document}

\title{Multi-field inflation from  a higher derivative $N=1$ Supergravity model  }

\author{
\vspace*{8mm}
{\bf G. A.~\ Diamandis} {\footnote{email: gdiam@phys.uoa.gr}}, \, {\bf B. C.~\ Georgalas}{\footnote{email: vgeorgal@phys.uoa.gr}}, \,  {\bf K.~\ Kaskavelis}{\footnote{email: kkaskavelis@phys.uoa.gr}}, \,{\bf  G.~\ Pavlopoulos}{\footnote{email: gepavlo@phys.uoa.gr}}}
\affiliation{National and Kapodistrian University of Athens, Department of Physics,\\
Nuclear and Particle Physics Section, GR--157 71  Athens, Greece}

\vspace*{2cm}  
\begin{abstract}
We study the cosmological evolution of a $N=1$ supergravity model, dual to a higher derivative supergravity model coupled to scalar fields so that in the Einstein frame the model is ghost free. We find that this model admit slow roll inflationary solutions showing essentially two-field inflation. The cosmological parameters calculated for this model lie in acceptable range with corrections of the order $lnN/N$, N being the number of e-foldings, compared to the Starobinsky inflation model.
\end{abstract}
\maketitle

\section{Introduction}
 The study of gravity theories including higher order terms of the curvature and its derivatives is well motivated since in the effective theory of gravity such terms are present \cite{Buchdahl, tHooft:1974toh, stelle1978}. The treatment of these theories is for many purposes  facilitated by the fact that they admit a dual description in terms of  Einstein gravity coupled to a certain number of scalar fields \cite{whitt, Wands, SOTIRIOU, CAPO}. The above property and the corresponding dual description has been extended to $N=1$ supergravity models \cite{THEISEN, CECOTTI}. The major problem for the use of the dual formulation in the study of these higher order theories is that the scalar fields necessarily introduced do not correspond to physical degrees of freedom leading thus to ghost states. This happens inevitably departing from quadratic terms in the curvature tensor and even in special cases of quadratic generalization of the Einstein gravity. Several attempts to deal with this problem and to construct ghost free models have been presented in the bibliography \cite{GHOST, BISWAS1, BISWAS2}. Certainly the study of the dual description in an attempt to understand the effect of higher order terms coming from quantum corrections has to do with the short distance behaviour of gravity. Interestingly enough the extra mode arising in a $R + R^2$ theory and in its supersymmetric extension seem to be relevant for  cosmology since in the dual description one of the corresponding scalar fields is responsible for the inflationary behaviour as is recognized in the celebrated Starobinsky's model. The above reasons justify both the history  which goes many years back and the recent revival of the study of generalizations of the Einstein (super)gravity.
 
In a relatively recent work a $N=1$ supergravity model has been presented with highest order terms of the form $R^3$ and $R \Box R$. Allowing kinetic terms for the chiral multiplets needed to be introduced in the higher order description the auxiliary fields become dynamical. In this way the physical degrees of freedom match,  leading to a ghost free dual form of $N=1$ Einstein supergravity coupled to four chiral multiplets. Furhermore a preliminary discussion of the cosmology of this model was performed, resulting to deformation of the Starobinky's model \cite{STAR}. In particular the potential has directions reminiscent of the Starobinsky model but in  the cosmological evolution at least two fields are relevant. Modifications of the Starobinsky inflation from generalized gravity models are also addressed in the literature \cite{Cuzinatto2018}. In this work we perform a detailed analysis of this model with the following order. In the next section we give a very short review of the model in discussion. In Sec. III we perform the analysis of the cosmological evolution pointing out that from the eight real scalars involved, only two are essential for the evolution, while six of them relax very quickly to their minimum value. In Sec. IV we calculate the basic cosmological parameters following the literature on multifield inflation \cite{Mukhanov1990,Seery2005, Kim2006, Kaiser2012,Kaneda2015}. In particular  corrections of the order $lnN/N$, $N$ denoting the number of e-foldings, to the Starobinsky model are found keeping this generalization viable. We conclude with the discussion in Sec. V. Many of the details of our analysis are presented in the two Appendices that close the work.

\section{Setting the model}
	The model under consideration is analysed in its basic characteristics in
	 \cite{Diamandis2017}. It is a higher $R$ supergravity model described by the following functions of chiral superfields,	 
\bea
\Omega&=&T+\bar{T}+(Q \bar{\Phi}+\Phi \bar{Q})+\omega(X,\bar{X},\Phi,\bar{\Phi},C,\bar{C},Q,\bar{Q}) \nonumber\\
W&=&T\Phi+QC+h(X,\Phi,C,Q) \label{omw}
\eea
with the specific choices for $\omega,h$,
\bea
\omega&=&2\alpha\, C \bar{C}+2\lambda\, Q\bar{Q}+2\beta \, \Phi\bar{\Phi} \nonumber \\
h(X,\Phi,C,Q)&=&h(\Phi,C)=\Phi f(C). \label{omw2}
\eea
where $\alpha, \, \lambda, \, \beta,$ are assumed to be positive.\\
The pure gravitational part in the higher $R$ description is given by
\be
e^{-1}L=-\frac{R}{3}f\left(\frac{R}{6}\right)+\frac{\alpha}{18}\left( \frac{R^3}{6}+R \Box R  \right)+ \frac{\beta}{18}R^2.
\ee
It is known that  $R^n$ supergravity models admit a dual description as Einstein gravity models which have ghost states for $n>2$ \cite{CECOTTI}.  
In the work mentioned above it was proven  that no ghost-states appear in the ordinary $N=1$ supergravity description at the cost of keeping coupling of certain scalar modes  in the higher $R$ description equalizing thus the physical degrees of freedom in the two descriptions. This was achieved via the introduction of the $ Q\bar{Q}$ and the $ C \bar{C}$ terms in the function $\omega$.

At the Einstein frame of $N=1$ supergravity description of the above model the action  is determined by the K\"ahler function
\be
\mathcal{G}=-3ln\lt(-\fr{\Omega}{3}\rt)+ln(W\bar{W})
\ee
and if we name the fields $T,Q,\Phi,C$ collectively as $\phi^I$ with $I=1,...4$ the action is given by
\be
S_{Einstein}=\int{d^4x\mathcal{ L}}=\int{d^4x \sqrt{-g}\lt[\fr{R}{2}-\fr{1}{2}\mathcal{G}_{I \bar{J} }g_{\mu\nu} \pt^\mu \phi^I \pt^\nu \bar{\phi}^{\bar{J}}-V(\phi^I,\bar{\phi}^{\bar{I}})\rt]} \label{ein1}
\ee
where $g_{\mu\nu}$ is the spacetime metric, $g$ its determinant, $R$ the Ricci scalar and
\be
\mathcal{G}_{I \bar{J} }=\pt_{\phi^{I}\bar{\phi}^{\bar{J}} }\mathcal{G},
\ee
the metric in the field space with the indices \textbf{$I,J$} running over the four complex scalar fields.

The potential is given by 
\be
V=e^{\mathcal{G}} \lt(\mathcal{G}^{\bar{I}J}\bar{\mathcal{G}}_{\bar{I}}\mathcal{G}_J-3\rt), \label{potdef}\\
\ee
where
$$\mathcal{G}^{\bar{I}J}=\mathcal{G}_{I\bar{J}}^{-1}, \quad \quad \mathcal{G}_I=\pt_{\phi^I}\mathcal{G} \;, \;\bar{\mathcal{G}}_{\bar{I}}=\pt_{\bar{\phi}^{\bar{I}}}\mathcal{G}.$$

 Furthermore, in that work it has been showed that the potential becomes stable with the addition of a stabilizer term $-\zeta |\Phi|^4$ proposed in \cite{Kallosh2013} to the function $\Omega$ and provided that the function $f(C)$ is at most quadratic in $C,$ of the form 
\bea
f(C)=f_0+f_1 C+f_2 C^2 \label{fc} 
\eea
with $f_0,f_1,f_2$ real coefficients{\footnote{
Consideration of more general functions of $C$ may be implemented  by introducing additional stabilizing term e.g. $-\zeta^{\prime} |C|^4$ but such an option is not examined  in this work.
}}. Then the potential has a global minimum with vanishing vacuum energy at $T=-f_0,\, \Phi=Q=C=0$.
Note also that the following resrtictions
\bea
f_0>0,\,4\beta\lambda-1>0,\, (\lambda f_1^2-f_1+4\alpha \lambda f_0+\beta)^2>4\alpha(4\beta \lambda -1)f_0.
\eea
ensures the absence of tachyonic states.

\section{Cosmological evolution}

In order to study  the cosmological behaviour of the model resulting from (\ref{ein1}), with the inclusion in $\omega$ of the term $-\zeta |\Phi|^4$ and $f(C)$ given by (\ref{fc}),  we assume a Friedmann-Robertson-Walker metric 
\bea
ds^2=-dt^2+a^2(t)\lt[dr^2+r^2 d\Omega\rt],
\eea
and  separating the fields $\phi^I$ into real and imaginary parts, we reparametrise our model in terms of eight real fields denoted by $z^I$ with $I=1,...,8$.   The field equations read
\bea
\ddot{z}^I+\Gamma^I_{JK}\dot{z}^J\dot{z}^K+3H\dot{z}^I+G^{IJ}V_{,J}=0, \label{cosm1}
\eea
where $G_{IJ}=\pt_{z^I z^J}\mathcal{G},G^{IJ}=\lt(G^{-1}\rt)_{IJ}\;, \; \Gamma^I_{JK}=\fr{1}{2}G^{IL}\lt(\fr{\pt G_{LJ}}{\pt z^K}+\fr{\pt G_{LK}}{\pt z^J}-\fr{\pt G_{JK}}{\pt z^L}\rt)$ and  the Hubble parameter is given by
\bea
H=\fr{\dot{a}}{a}=\fr{1}{\sqrt{3}}\sqrt{V+\fr{1}{2}G_{IJ}\dot{z}^I\dot{z}^J}.
\eea
The fields relevant for the study of the cosmology are  $w^i=\{ReT,ReC\}$, while the remaining  six fields $\chi^i=\{ImT,ReQ,ImQ,ImC,Re\Phi,Im\Phi\}$ will be considered to relax to their zero minimum value. In particular in the Appendix A it is shown that these fields tend rapidly to their minimum value leaving the fields $w^i$ to drive the cosmological evolution of the model. Being restricted in the space of these two fields the potential reads

\bea
V( ReT, ReC ) \, &=& \, \dfrac{ 9 }{ 8 \lambda \, ( 4 \beta \lambda - 1 \,) \, ( ReT + \alpha  ReC^2   )^2  } \, 
\nonumber \\ 
& &
  \cdot \,\Big( \, ( 4 \beta \lambda - 1 \,) \, ReC^{\,2} + \left( ReC - 2 \lambda \, ( ReT +f(ReC) )\right) ^{\,2} 
  \Big). \nonumber \\
\label{pot1}  
\eea
which is a positive semidefinite function  and reparametrising as $ReT=-f_0 e^{\sqrt{\fr{2}{3}}\psi}$ and $ReC\equiv c $   the corresponding metric becomes \footnote{It is easily checked that either setting the fields $\chi^i$ to their minimum value in the set of the full equations in  (\ref{cosm1}), or working with the restricted lagrangian, leads to the same set of equations to be studied. }

\bea
G_{\psi\psi}\,=\, \frac{b^2}{X^2}, \quad G_{\psi c} \,= \, -\frac{2\sqrt{6}\alpha b c}{X^2}, \quad G_{c c} \,=\, \frac{12\alpha b}{X^2}, \quad \text{where} \quad b\,=\,e^{\sqrt{\frac{2}{3}}\psi}, \quad X \,=\, b-2 \alpha c^2.
\eea
The metric can be brought into diagonal form  if we set \footnote{From now on we take $f_0 = 1/2$  which is necessary in order to have canonically normalized Einstein  gravity in the dual description, see \cite{Diamandis2017}}.

\bea
x &\equiv& \frac{1}{e^{\sqrt{\frac{2}{3}} \psi }-2 \alpha c^2}\nonumber \\
y &\equiv& \fr{c}{\frac {2 f_ 1\lambda - 1} {4\lambda\left (\alpha - f_ 2 \right)}}. \label{xydef}
\eea

 The choice of the constant $c_{as}\equiv \frac {2 f_ 1\lambda - 1} {4\lambda\left (\alpha - f_ 2 \right)}$ in the normalization of $c$ will be explained later on. In the above basis the  metric elements  become 

\bea
G_{xx}&=&\frac{3}{2 x^2}\: , \: G_{xy}=0 \:,\: G_{yy}=12 \alpha c_{as}^2\,  x .
\eea

and the potential gets the form

\bea
V(x,y)=V_0+V_1(y)x+V_2(y)x^2
\eea

with 

\bea
V_0&=&\frac{9 \lambda }{2 (4\beta  \lambda -1)} \nonumber \\   
V_1&=& \left[-\left(1+\frac{\left(1-2 f_1 \lambda \right){}^2}{8 \lambda ^2 \left(\alpha -f_2\right)}\right)+\frac{(y-1)^2 \left(1-2 f_1 \lambda \right){}^2}{8 \lambda ^2 \left(\alpha -f_2\right)} \right]\,V_0 \nonumber \\
V_2&=&V_2^{(0)}+V_2^{(1)}(y-1)+V_2^{(2)}(y-1)^2+V_2^{(4)}(y-1)^4 \label{potxy}
\eea

with the constant $V_2^{(4)}>0$.

As is already mentioned the potential is positive apart from the global minimum at $ReT=-\fr{1}{2},ReC=0$ or equivalently at $x=1,y=0$ and possibly  along the direction $y=\tilde{k}\fr{1}{\sqrt{x}}$ and the limit $x\rightarrow 0$, where the potential tends to the value 

\bea
V(x,y=\tilde{k}\fr{1}{\sqrt{x}})\stackrel{x\rightarrow 0}{\rightarrow}\frac{9 \lambda  \left(\frac{\tilde{k}^2 \left(1-2 f_1 \lambda \right){}^2}{8 \lambda ^2 \left(\alpha -f_2\right)}+1\right){}^2}{2 (4 \beta \lambda -1)} \label{potlim}
\eea

and by demanding it to be zero, it has real solutions for $\tilde{k}$ only if $\alpha-f_2<0$. In order to exclude this case we shall take from now on $\alpha>f_2$ \footnote{An inflation scenario of hill-top form is not considered in this work. }. As we will see later for large number of efoldings $N$ the field $y$ takes the value $y\approx 1$ or equivalently $c\approx c_{as}$ during inflation. The fine tunning case $\alpha=f_2$ means there is no asymptote value for $y$ or $c$ during inflation and this case needs special treatment which we have not done in this work. 

\subsection{Evolution of the fields}
The behaviour of the solutions is understood analytically although the full solution can be obtained only numerically. 
 Since we are interested for trajectories initiating away from the minimum of the potential we consider the equations for small $x$ (large negative $ReT$).  The equations of motion for the fields $\tilde{\varphi}^i=\{x,y\}$  at lowest order in $x$ read in this case:

\bea
\ddot{\tilde{\varphi}}^i+\Gamma^i_{jk}\dot{\tilde{\varphi}}^j\dot{\tilde{\varphi}}^k+3H\dot{\tilde{\varphi}}^i+G^{ij}V_j&=&0\Rightarrow\nonumber \\
\ddot{x}-\fr{\dot{x}^2}{x}+3 \sqrt{\fr{3}{2}}  \sqrt{\frac{\lambda }{4 \beta  \lambda -1}}\dot{x}-4\alpha c_{as}^2x^2\dot{y}^2-W x^2+Y x^2(y-1)^2&=&0 \nonumber \\
\ddot{y}+\fr{\dot{x}\dot{y}}{x}+3 \sqrt{\frac{3}{2}} \sqrt{\frac{\lambda }{4 \beta  \lambda -1}}\dot{y}+\frac{3 \lambda  (y-1) (\alpha-f_2 )}{\alpha  (4 \beta  \lambda -1)} \,+\, \frac{3x}{4 \lambda \alpha}&=&0, \label{eqas}
\eea
with the constants $W,Y$ being positive and $\Gamma^i_{jk}$ denote now the Christoffel symbols in the space of the fields $\tilde{\varphi}^i$. The equations for $x$ and $y$ are written in the lowest non-linear approximation.

Adopting the iteration method for the solution of the non-linear equations  and starting from the linear part  of the $x$ equation in eq.(\ref{eqas}), that is from the three first term we see that they give a solution for $x$ which has a negligible contribution to the $y$ equation above. So to zero order in $x$ we get:

\bea
\ddot{y}&+&3 \sqrt{\frac{3}{2}} \sqrt{\frac{\lambda }{4 \beta  \lambda -1}}\dot{y}+\frac{3 \lambda  (y-1) (\alpha-f_2 )}{\alpha  (4 \beta  \lambda -1)}=0\Rightarrow \nonumber \\
y-1&=&C_1 \exp \left(-\frac{1}{2} \sqrt{\frac{3}{2}} t \sqrt{\frac{\lambda }{4 \beta  \lambda -1}} \left(3+\frac{\sqrt{\alpha +8 f_2}}{\sqrt{\alpha }}\right)\right)+\nonumber \\
&+&C_2 \exp \left(-\frac{1}{2} \sqrt{\frac{3}{2}} t
   \sqrt{\frac{\lambda }{4 \beta  \lambda -1}} \left(3-\frac{\sqrt{\alpha +8 f_2}}{\sqrt{\alpha }}\right)\right). \label{yas}
\eea

Clearly the quantity $3-\frac{\sqrt{\alpha +8 f_2}}{\sqrt{\alpha }}>0$ as $\alpha>f_2$ and therefore both exponents are negative \footnote{Note that if we set $C_2=0$ we may have an acceptable hill-top scenario even for $\alpha < f_2.$}. Consequently, the field $y$ settles down  to values near  $y=1$ for small $x$. Note that the factor $\frac {2 f_ 1\lambda - 1} {4\lambda\left (\alpha - f_ 2 \right)}$ in the definition of $y$ in (\ref{xydef}) is chosen  so that $y=1$ becomes the asymptotic value for the minimum of the potential in the $y$ direction for small $x$. In the following Figure (\ref{fig1}) we show the evolution for the fields $x,y$ for some representative values of the parameters of the potential and different initial values for $x$.

\begin{figure}
\centering
\begin{minipage}{.5\textwidth}
  \centering
  \includegraphics[width=.8\linewidth]{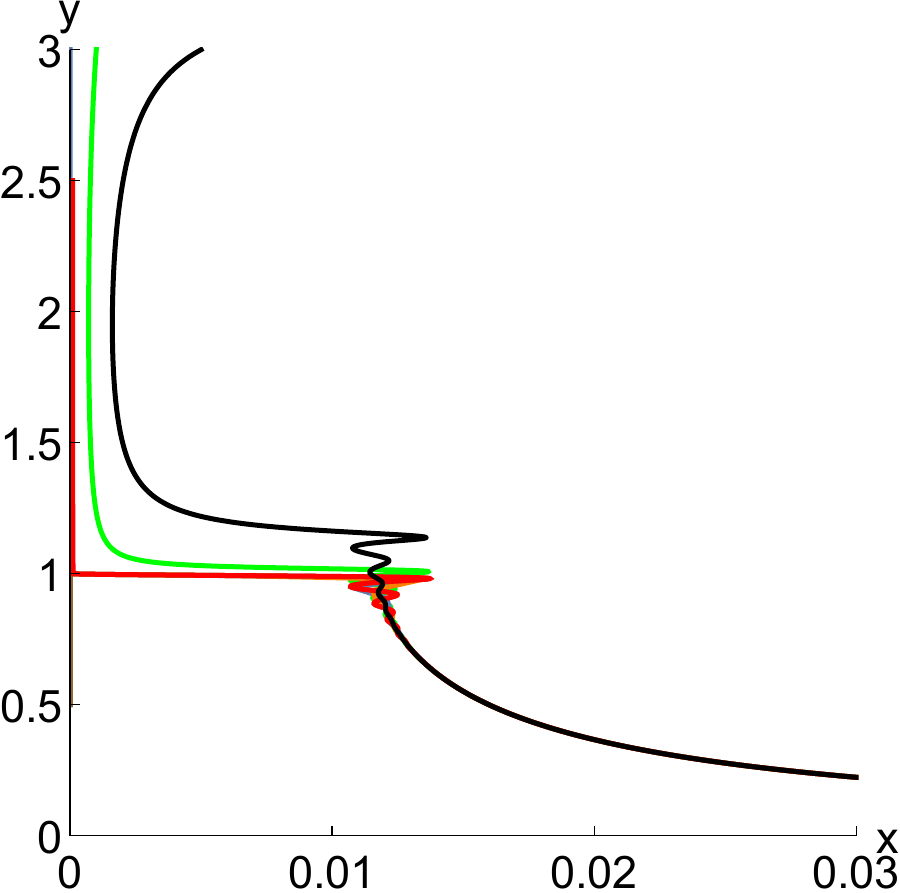}  
\end{minipage}%
\begin{minipage}{.5\textwidth}
  \centering
  \includegraphics[width=.8\linewidth]{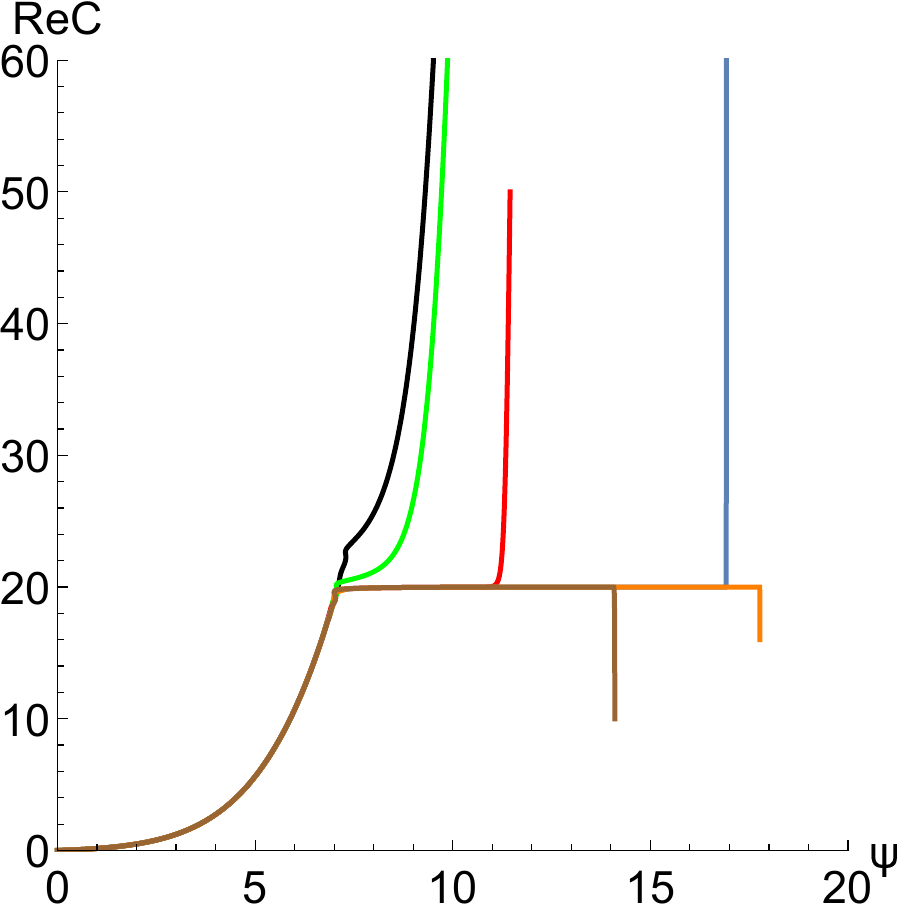}
\end{minipage}
\caption{Left panel : The evolution of the values of the fields $x,y$ with initial values $x=5\cdot 10^{-7},y=0.8$ ( Orange Line ), $x=10^{-6},y=3$ ( Blue Line ), $x=10^{-5},y=0.5$ ( Brown Line ), $x=10^{-4},y=3$ ( Red Line ), $x=10^{-3},y=3$ ( Green Line ), $x=5\cdot 10^{-3},y=3$ ( Black Line ). The values of the parameters of the potential are $\alpha=0.3, \beta=0.5, \lambda=1.5, f_1=\fr{13}{3},f_2=0.2$.  
Right panel: The evolution of the values of the fields $\psi,ReC$ with the same initial values and parameters of the potential as before but in the $\psi,ReC$ basis, namely $\psi=17.77, ReC=0.8 c_{as}=16$ ( Orange Line ), $\psi=16.9, ReC=3c_{as}=60$ ( Blue Line ), $\psi=14.1, ReC=0.5 c_{as}=10$ ( Brown Line ) , $\psi=11.5, ReC=3c_{as}=60$ ( Red Line ), $\psi=9.5, ReC=3c_{as}=60$ ( Green Line ) and 
$\psi=9.8, ReC=3c_{as}=60$ ( Black Line ) .  }
\label{fig1}
\end{figure}

The numerical results confirm the above analytical consideration. The field $y$ is driven towards the minimum in the $y$ direction which for very small $x$ is at $y=1$. This indicates that in 
 the $(x,y)$ plane there is a trajectory perpendicular to which the potential is convex and forces the fields $x,y$ to follow this trajectory.
In order to find the trajectory $y(x)$ which the fields $x,y$ follow after they stabilize their motion 
we use the Hamilton-Jacobi equation, derived from the field equations, 
\bea
2\lt(G^{xx}H_x^2+G^{yy}H_y^2\rt)=3H^2-V.
\eea
Expanding it in powers of $x$  according to the Frobenius-Fuchs method
\bea
H(x,y)=x^s\sum_{n=0}^{\infty}{H_n(y)x^n}. \label{frob}
\eea
we find that the starting power $s$ of the expansion (\ref{frob}) must be $s=0$ or $s=-\fr{3}{2}$.
However, for a trajectory that starts at $x<<1$, for $s=-\fr{3}{2}$ the kinetic term $\fr{1}{2}\lt(G_{xx}\dot{x}^2+G_{yy}\dot{y}^2\rt)=2\lt(G^{xx}H_x^2+G^{yy}H_y^2\rt)$ is much larger than $V(x,y)\rightarrow V_0$ (the kinetic term starts in an expansion in $x$ with a negative power of $x$, whereas the potential with a constant)  and therefore it seems that  we don't have a slow roll inflation which demands $\fr{1}{2}\lt(G_{xx}\dot{x}^2+G_{yy}\dot{y}^2\rt)<<V(x,y)$. Nevertheless the numerical solution shows that even this case leads, after a few e-foldings, to slow roll evolution although it is not easy to be proven analytically.

If we restrict  the expansion (\ref{frob}) to $s=0$ which obviously satisfy the slow-roll condition we obtain an ordinary power series. 

Inserting the expansion in (\ref{frob}) in the Hamilton-Jacobi
equation and furthermore expanding the functions $H_i(y),H_i'(y)$ involved  in  series in $y$ around $y=1$, which is an attractor for $x<<1$, we can calculate the quantities $H_i(1),H_i'(1),H_i''(1),H_i'''(1),\ldots$.
 Furthermore expanding $y$ as 
\bea
y=1+y_1 x+y_2 x^2+\ldots \label{yansatz}
\eea
 we can  solve the equation of the trajectory $y(x)$ 

\bea
\fr{dy}{dx}=y_1+2 y_2 x+\ldots=\fr{G^{yy}H_y}{G^{xx}H_x}
\eea
using the coefficients $H_i(1),H_i'(1),H_i''(1),H_i'''(1),\ldots$ that we have calculated from the Hamilton-Jacobi equation.

The result is

\bea
y_1&=& -\frac{4 \beta  \lambda -1}{4 \lambda ^2 \left(\alpha -f_2\right)} \nonumber \\
y_2&=& \frac{(4 \beta  \lambda -1) \left(4 \beta  \lambda-1 +4 \lambda ^2 \left(\alpha +f_2\right) \left(\frac{(1-2 f_1 \lambda )^2}{8 \lambda ^2 \left(\alpha -f_2\right)}+1\right)\right)}{16 \lambda ^4 \left(\alpha-f_2\right){}^2}. \label{y1y2}
\eea

and for the velocities $\dot{x},\dot{y}$ for small $x$:

\bea
\dot{x}&=&v_2 x^2+v_3 x^3 \nonumber \\
\dot{y}&=&y_1 v_2 x^2+\lt(2v_2 y_2+v_3 y_1\rt)x^3 \label{dotxdoty}
\eea

with 

\bea
v_2&=&2 \sqrt{\frac{2}{3}} \sqrt{\frac{\lambda }{4 \beta \lambda -1}} \left(\frac{(1-2 f_1 \lambda )^2}{8 \lambda ^2 (\alpha-f_2)}+1\right) \nonumber \\
v_3&=&-\frac{\left(8 \lambda ^2 (\alpha-f_2)+(1-2 f_1 \lambda )^2\right)^2+9 (4 \beta \lambda -1) (1-2 f_1 \lambda )^2}{36 \sqrt{6} \lambda ^{7/2} (\alpha-f_2)^2 \sqrt{4 \beta \lambda -1}}.
\eea

The form of this trajectory is also confirmed from the numerical solutions with the mere assumption that the initial conditions imposed are far from the minimum of the potential as is the case for the cosmological study of the model.

\section{Calculation of the number of e-foldings and of cosmological parameters}
In this subsection we summarize the results concerning the cosmological parameters leaving the details which are presented in Appendix B. 
 We will use the  trajectory found in the previous subsection as it is an attractor for all the trajectories that start at asymptotic values of  $x$ and $y$. This is necessary in order to have the required  high number of e-foldings. The number of e-foldings with the redefinitions $$L\equiv 4\beta \lambda-1, \rho  \equiv \alpha-f_2$$ and (the already defined) $$c_{as} \equiv \frac {2 f_ 1\lambda - 1} {4\lambda\left (\alpha - f_ 2 \right)}$$ is given by:

\bea
N&=&\int{dN}=\int{H(x,y)dt}=-\int{\fr{1}{\epsilon}dlnH}= \nonumber \\
&=&-\int_{x_{\ast},y_{\ast}}^{x_{end},y_{end}}{\fr{1}{\epsilon H}\lt(\fr{\pt H}{\pt x}dx+\fr{\pt H}{\pt y}dy\rt)}\Big|_{y=1+y_1 x+y_2 x^2+\ldots}=\nonumber \\
&=&\frac{3 \left(\frac{1}{x_*}-\frac{1}{x_{end}}\right)}{4(1+2 \rho c_{as}^2)}+\log \left(\frac{x_{end}}{x_*}\right) \left(\frac{12 \beta ^2 L c_{as}^2}{(L+1)^2 \left(1+2 \rho c_{as}^2\right){}^2}-\frac{5}{12}\right)+\mathcal{O}(x_*^0)=\nonumber \\
&=&\frac{3 \left(\frac{1}{x_*}-\frac{1}{x_{end}}\right)}{4 A}+\log \left(\frac{x_{end}}{x_*}\right) \left(\fr{12B}{A^2}-\frac{5}{12}\right)+\mathcal{O}(x_*^1,x_{end}^1), \label{efolds}
\eea

with $\epsilon\equiv -\fr{\dot{H}}{H^2}$, $$A=1+2\rho c_{as}^2,B=\frac{\beta ^2 L c_{as}^2}{(L+1)^2}.$$  The values $x_*,y_*$ are the pivot values of the fields $x,y$ and $x_{end},y_{end}$ are the values of the fields $x,y$ where inflation ends.

For later use we  can  also   express $x_*$ as a function of $N$ using the expansion for large $N$: 

\bea
x_*=A_1 \fr{1}{N} +A_2 \fr{1}{N^2} lnN+A_3 \fr{1}{N^2}+\ldots \label{xastN}.
\eea

Substituting it in (\ref{efolds}) and equating the equal powers of $N$ we get

\bea
A_1&=&\fr{3}{4(1+2\rho c_{as}^2)}=\fr{3}{4A}, \nonumber \\
A_2&=&\frac{ \left(5-\frac{144 \beta ^2 L c_{as}^2}{(L+1)^2 \left(2 \rho c_{as}^2+1\right){}^2}\right)}{16\left(1+2 \rho c_{as}^2\right)}=\fr{\lt(5-\fr{144B}{A^2}\rt)}{16A} \nonumber \\
A_3&=&\frac{\left(144 B-5 A^2\right) x_{end} \log \left(\frac{4 A x_{end}}{3}\right)-9 A}{16 A^3 x_{end}}+\mathcal{O}(x_{end}^1).\label{xNcoef}.
\eea

The slow roll matrix $\epsilon^{IJ}$ is given by \cite{Seery2005}:

\bea
\epsilon^{IJ}=\fr{\dot{\phi}^I\dot{\phi}^J}{2H^2}=2G^{IM}G^{JN}\fr{H_{,M}H_{,N}}{H^2}.
\eea

If we are restricted in the fields $\tilde{\varphi}^i$,  after the substitution $y=1+y_1 x+y_2 x^2+\ldots$ all elements are only functions of $x$ and are of lowest order $x^4$ i.e. $\epsilon^{IJ}=\mathcal{O}(x^4)$ and are therefore small for small $x$.
The slow roll parameter $\epsilon_H$ is defined as $\epsilon_H=-\fr{\dot{H}}{H^2}=G_{IJ}\epsilon^{IJ}$ and comes out to be with the redefinitions :

\bea
\epsilon_H&=&\frac{4}{3} x^2 \left(1+2 \rho c_{as}^2\right){}^2+\frac{8}{27} x^3 \left(1+2 \rho c_{as}^2\right) \left(5 \left(1+2 \rho c_{as}^2\right){}^2-\frac{144 \beta ^2 L c_{as}^2}{(L+1)^2}\right)= \nonumber \\
&=&\frac{4}{3} x^2 A^2+\frac{8}{27} x^3 A \left(5 A^2-144 B\right),
\eea
being also small for $x<<1$ where inflation occurs.

As far as the spectral index $n_s$ and the tensor to scalar ratio $r$ are concerned we remind that  in our case we have eight real scalars, and although six of them take quickly the zero value, the perturbations of the metric and the fields in these directions have to be taken into account, as in principle they may affect the cosmological observables. In Appendix \ref{app1} we prove that they don't play any role in the cosmological observables $n_s,r$. In Appendix \ref{app2} we repeat the formulation described mainly in  \cite{Lalak} and \cite{vandeBruck} for the calculation of  $n_s,r$ for the case of two field inflation.

\subsection{Spectral index}

The spectral index is defined in terms of the power spectrum $ \mathcal{P}_\mathcal{R}$  by 

\bea
n_s-1=\fr{dln \mathcal{P}_\mathcal{R}}{dlnk}
\eea

at $k=aH$. Note that $\fr{dlnk}{dt}=\fr{dln(aH)}{dt}=\fr{\dot{a}}{a}+\fr{\dot{H}}{H}=(1-\epsilon)H$
so 
\bea
\fr{d}{dlnk}=\fr{1}{1-\epsilon}\fr{1}{H}\fr{d}{dt}.
\eea 

The result, keeping the leading and subleading term, with $P_{\mathcal{R}_*}$ calculated in Appendix \ref{app2} in (\ref{pr2}) and using (\ref{dotxdoty}) for the time derivative of $x$, is:

\bea
n_s&=&-\frac{8}{3} x_{*} \left(2 \rho c_{as}^2+1\right)+\frac{8}{27} x_{*}^2 \left((12 K-19) \left(2 \rho c_{as}^2+1\right){}^2+\frac{288 \beta ^2 L c_{as}^2}{(L+1)^2}\right).
\eea

With the substitutions  $A\equiv 1+2\rho c_{as}^2, \,B\equiv\frac{\beta ^2 L c_{as}^2}{(L+1)^2}$ and with $ K=2-ln2-\gamma\approx 0.7296$ (with $\gamma$ the Euler-Mascheroni constant) for $n_s$ reads: 

\bea
n_s=-\frac{8 A x_{*}}{3}+\frac{8}{27} x_{*}^2 \left(\lt(12K-19\rt)A^2+288 B\right).
\eea

Substituting $x_{*}$ as a function of the number of e-foldings by (\ref{xastN}) we get 

\bea
n_s&=&-\fr{2}{N}+\fr{5A^2-144B}{6A^2}\fr{lnN}{N^2}+\nonumber \\
&+&\fr{1}{N^2}\lt(\frac{3}{2 A x_{end}}+\left(\frac{5}{6}-\frac{24 B}{A^2}\right) \log \left(\frac{4 A x_{end}}{3}\right)+\frac{48 B}{A^2}+2 K-\frac{19}{6}+\mathcal{O}(x_{end}^1)\rt).
\eea

\subsection{Tensor to scalar ratio}

The tensor power spectrum is given by \cite{Stewart1993}

\bea
\mathcal{P}_{\mathcal{T}_*}=8\lt(\fr{H_{*}}{2\pi}\rt)^2\lt(1+2(K-1)\epsilon_{*}\rt)
\eea

with $K=2-\ln2-\gamma$.

Keeping the leading and subleading term in $x$, we have

\bea
\mathcal{P}_{\mathcal{T}_*}=\frac{3 c_{as}}{4 \pi ^2 \sqrt{B L}}-\frac{3 A  c_{as}}{2 \pi ^2 \sqrt{B L}}x+\mathcal{O}(x^2)
\eea

and by taking the ratio $r=\fr{\mathcal{P}_{\mathcal{T}_*}}{\mathcal{P}_{\mathcal{R}_*}}$ with $\mathcal{P}_{\mathcal{R}_*}$ given by (\ref{pr2}) we get

\bea
r=\frac{64 A^2 x_{*}^2}{3}-\frac{128}{27} A^3 x_{*}^3 \left(\frac{144 B}{A^2}+12 K-5\right),
\eea

Substituting $x_{*}$ as a function of the number of e-foldings by (\ref{xastN}) the result reads

\bea
r&=&\fr{12}{N^2}-\fr{2(5A^2-144B)}{A^2}\fr{lnN}{N^3}+\nonumber \\
&+&\fr{1}{N^3} \lt( -\fr{18}{Ax_{end}}+\frac{2 \left(\left(144 B-5 A^2\right) \log \left(\frac{4 A x_{end}}{3}\right)+A^2 (12 K-5)+144 B\right)}{A^2}+\mathcal{O}(x_{end}^1)\rt).
\eea

For the determination of $x_{end}$ we require the absolute value of $\eta_{\sigma\sigma}$ (which is defined as $\eta_{\sigma\sigma}\equiv\frac{V_{\sigma\sigma}}{3H^2} $ with $V_{\sigma\sigma}$ defined in (\ref{Vssdef})) to be smaller than unity, therefore \footnote{The results are not sensitive in different way of determining  $x_{end}$.}

\bea
\eta_{\sigma\sigma}&=&-\frac{4 A x_{end}}{3}+\frac{128 B x_{end}^2}{3}=-1 \label{end1}\Rightarrow
x_{end}=\fr{3}{2A}\fr{1}{1+\sqrt{1-96\Lambda}}
\eea

with $\Lambda=\fr{B}{A^2}.$ The eq. (\ref{end1}) has a solution for $\Lambda<\fr{1}{96}.$  For $\Lambda>\fr{1}{96}$  we require as an estimation for where inflation ends

\bea
\fr{d\eta_{\sigma\sigma}}{dx}\Big|_{x=x_{end}}=0\Rightarrow x_{end}=\fr{A}{64B}=\fr{1}{64\Lambda A }.
\eea

Then the cosmological parameter $n_s$ can be written for $\Lambda<\fr{1}{96}$:
 
\bea
n_s&=&-\fr{2}{N}+\fr{5-144\Lambda}{6}\fr{lnN}{N^2}+\nonumber \\
&+&\frac{12 K+288 \Lambda +6 \sqrt{1-96 \Lambda }+(5-144 \Lambda ) \log \left(\frac{2}{\sqrt{1-96 \Lambda }+1}\right)-13}{6 N^2} \label{ns1}
\eea

and for $\Lambda>\fr{1}{96}$

\bea
n_s&=&-\fr{2}{N}+\fr{5-144\Lambda}{6}\fr{lnN}{N^2}+\nonumber \\
&+&\frac{12 K+864 \Lambda +(5-144 \Lambda ) \log \left(\frac{1}{48 \Lambda }\right)-19}{6 N^2}. \label{ns2}
\eea

The tensor to scalar ratio can be written for $\Lambda<\fr{1}{96}$

\bea
r&=&\fr{12}{N^2}-2(5-144\Lambda)\fr{lnN}{N^3}+\nonumber \\
&+&\frac{2 \left(12 K+144 \Lambda -6 \sqrt{1-96 \Lambda }+(144 \Lambda -5) \log \left(\frac{2}{\sqrt{1-96 \Lambda }+1}\right)-11\right)}{N^3} \label{r1}
\eea

and for $\Lambda>\fr{1}{96}$:

\bea
r&=&\fr{12}{N^2}-2(5-144\Lambda)\fr{lnN}{N^3}+\nonumber \\
&+&\frac{2 \left(12 K-432 \Lambda +(144 \Lambda -5) \log \left(\frac{1}{48 \Lambda }\right)-5\right)}{N^3}. \label{r2}
\eea

\begin{figure}[h]
\centering
  \includegraphics[width=.8\linewidth]{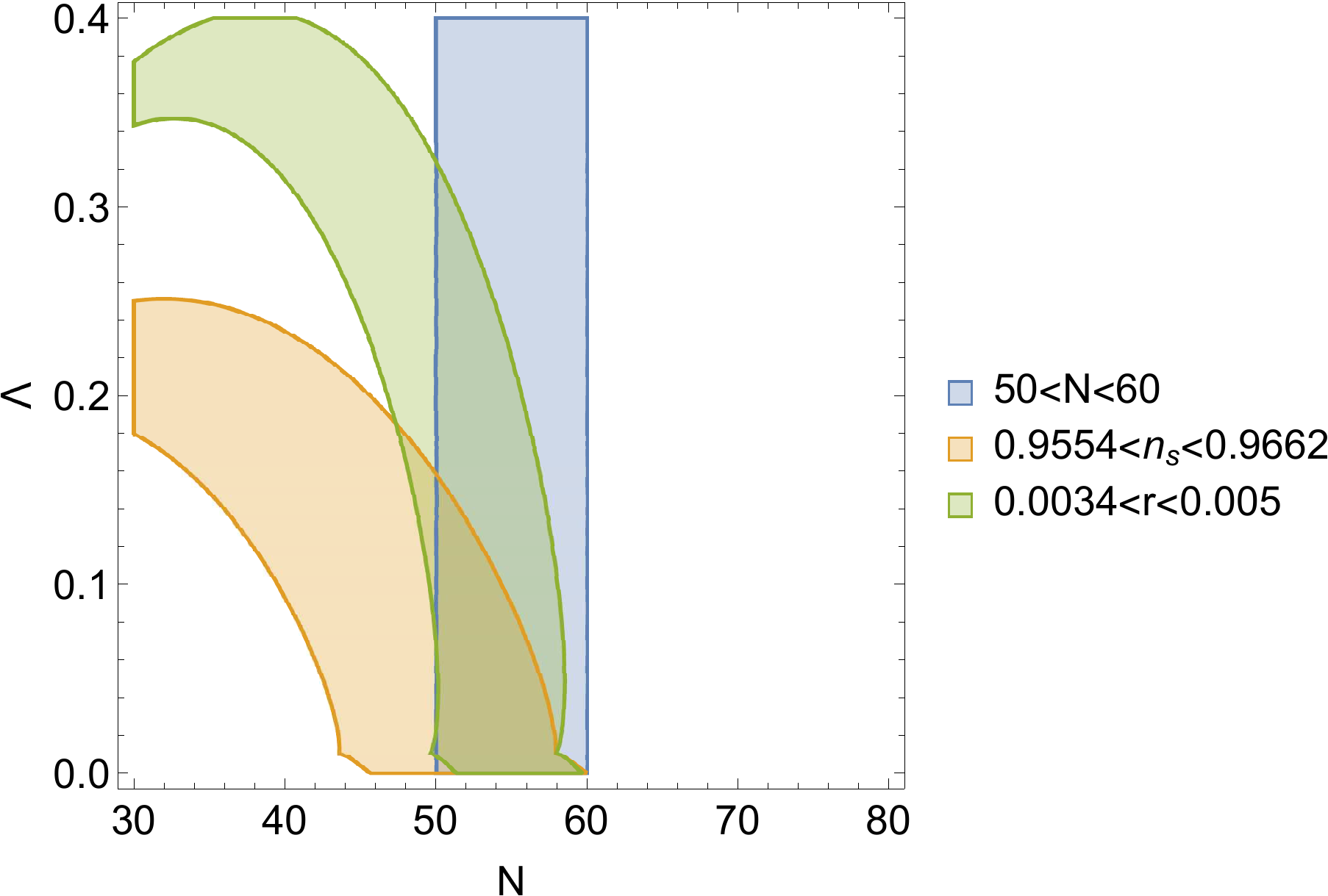}  
\caption{ The physical region is where the three individual regions overlap. That is the number of e-foldings is between $50$ and $60$ and $n_s$ is within the experimental bounds. Furthermore, we observe that the appropriate values of $r$ are those of the Starobinsky model. We see that the overlapping ends when $\Lambda>0.16.$}
\label{fig2}
\end{figure}

In figure (\ref{fig2}) we plot $n_s,r$ given by eqs (\ref{ns1})-(\ref{r2}) as a function of the number of e-foldings $N$ and of the parameter $\Lambda$ and we show indicatively the physical region dictated by observations \cite{Planck,WMAP,BICEP2, Ade2015} for $n_s,r$ and the number of e-foldings $N$.

We observe that the maximum allowed value of $\Lambda$ is $\Lambda \approx 0.16$ and that the tensor to scalar ratio is small, $0.0034<r<0.005$, typical for the Starobinsky model. 
Concluding we have an in principle viable model giving small deviations from the original one field inflation Starobinsky model.

\section{Summary - Conclusions}

In this work we make a detailed study of the properties regarding the inflationary behaviour of a $N=1$ supergravity model. This model involves four chiral multilpets and it is dual to higher order supergravity containing $R^3$ and $R \Box R$ terms. These terms are purely gravitational and of $\kappa^2$ order, where $\kappa$  is the gravitational coupling constant. Despite the fact that there is a  qubic term in the scalar curvature, its dual description is ghost free. This happens since in the higher order description gravity remains coupled to certain complex scalar fields and notably to one of the auxiliary fields of the chiral multiplets maching thus the physical degrees of freedom between the two descriptions. This occurs by including in the Lagrangian kinetic terms for the chiral mutiplets which cease from being just Lagrange multipliers.

The basic features of the cosmology derived from this model are the following. From the eight scalar fields involved, only two are crucial for the cosmological evolution, under very mild and natural conditions. There are solutions exhibiting slow roll behaviour in which two fields take part while the rest relax almost immediately to their minimum value. The resulting evolution certainly suggests relation to the existence of attractors studied in the bibliography  \cite{Kallosh2013, Galante2014, Achucarro2017, Iarygina2018,  Christodoulidis20191, Christodoulidis2019}, although the precise connection is not explored in this work. Furthermore perturbations of these fields do not affect considerably the cosmological parameters. From the calculation of the cosmological parameters it turns out that the leading order correction to the result from the Starobinsky model, which is phenomenologically viable, is of order $ln N /N$ where N is the number of e-foldings. Therefore the generalization considered yield reasonable results regarding the slow roll inflation and the cosmological parameters evaluated from this.

Concluding two remarks are in order. The first has to do with the fact that the requirement for a stable potential led to consider the function $f(C)$ to be quadratic in $C$. For general $f(C)$ we expect that a term of the form $-\gamma \left(C\,\bar{C}\right)^2$  in the  K\"ahler function is adequate to stabilize the potential as is the case with the term introduced by Kallosh for the chiral field $\Phi$. Note that such terms exist in the effective K\"ahler function. The sign of the coefficient may lead to stability or further instability. The second remark concerns the scalar fields required to work in the dual description.  We have seen that only two of them are crucial for cosmology that is they have to do with the long range behaviour of the model. We expect that this feature holds if we consider even higher order terms of the effective action. That is,  although in the Einstein frame in fact a large  number of scalar fields are present, only few of them are crucial for cosmology and the majority have to do with the ultraviolet behaviour of the theory. Certainly definite answers in favour or against the above claims need much further work.

\vspace*{4mm}  
{\textbf{Acknowledgements}}   

This research has been financed by NKUA ( National Kapodistrian University of Athens ).  The authors   wish to thank A. B. Lahanas and V. Spanos  for illuminating discussions.

\appendix

\section{Determining  the degrees of freedom essential for cosmology.} \label{app1}

In this appendix we show that the cosmologically relevant fields are  the two fields $y^i\equiv\{ReT,ReC\}$ while the  six fields $ImT,ReQ,ImQ,ImC,Re\Phi,Im\Phi$ obtain very quickly zero value.
 If we linearize the system of equations (\ref{cosm1}) towards the six fields $\chi^i\equiv\{ImT,ReQ,ImQ,ImC,Re\Phi,Im\Phi\}$ we obtain  equations of the form

\bea
\ddot{\chi^i}=M^i_j(y^i,\dot{y}^i) \chi^j+N^i_j(y^i,\dot{y}^i) \dot{\chi}^j 
\eea

and defining $p_i\equiv\dot{\chi}^i$, we have the first order equations

\bea
\left( \begin{array}{c}
\dot{\chi}  \\
\dot{p}  \end{array} \right)=\left( \begin{array}{cc}
0 & 1 \\
M & N
  \end{array} \right)
	\left( \begin{array}{c}
\chi \\
p
  \end{array} \right)
\eea

The real parts of the eigenvalues of the matrix
$\mathcal{M}\equiv\left( \begin{array}{cc}
0 & 1 \\
M & N
  \end{array} \right)$ are negative for $ReT,ReC$ in the regime $x\approx 0, y\approx 1$,  with $x,y$ defined in (\ref{xydef}),  if we consider as  plausible values for the velocities $Re\dot{T},Re\dot{C}$ those given by the slow roll motion $3H \dot{y}^i=-G^{ij}V_j$ with $H=\sqrt{\fr{V}{3}}$ on the plane spanned by $y^1,y^2$ (all other fields equal to zero). Then the eigenvalues are 
\bea
\lambda_{1,2}&=&-\sqrt{6} \sqrt{\frac{\lambda }{4 \beta \lambda -1}}, \nonumber \\
\lambda_{3,4}&=&-\sqrt{3} \sqrt{\frac{\lambda }{8 \beta \lambda -2}}, \nonumber \\
\lambda_{5,6}&=&\frac{1}{2}  \sqrt{3} \left(\mp i\sqrt{7}-3 \right) \sqrt{\frac{\lambda }{8 \beta \lambda -2}}, \nonumber \\
\lambda_{7,8}&=&-\frac{\mp \sqrt{3} \left(\sqrt{\alpha \lambda  (\alpha-8 f_2)}+3 \alpha \sqrt{\lambda }\right)}{2 \alpha \sqrt{8 \beta \lambda -2}}, \nonumber \\
\lambda_{9,10;11,12}&=&-\frac{3}{2} \sqrt{3} \sqrt{\frac{\lambda }{8 \beta \lambda -2}} \mp \frac{2 i \sqrt{6} \sqrt{\zeta } \left(\frac{\lambda }{4 \beta \lambda -1}\right)^{3/2}}{\sqrt{x}}. 
\eea	
The existence of such eigenvalues at the linearized level guarantees, according to Picard's iteration method, an iterative solution of the full system with each term in the iteration procedure converging even faster than the linear one.

The numerical analysis indicates that the condition imposed is not necessary for the eigenvalues to have negative real parts although this is difficult to be shown analytically for the full case. Nevertheless it can be shown in a reduced case where $ImT, \,Re\Phi, \,Im\Phi$ are taken to be zero. Note that this is the case already in the supersymmetrization of the Starobinsky model. In this case the matrices $M$, $N$ are diagonal and as is can be easily seen negative definite with the mere condition adequate for the asbsence of tachyonic fields which is already imposed. The equation determining the eigenvalues of the reduced matrix is equivalent to solve three quadratic equations
\[
\lambda^2 \,-\, \lambda \, \nu_{i} \,-\, \mu_{i} \,=\, 0
\]
where $ \mu_i , \, \nu_{i}$ the negative diagonal elements of the matrices $M$, $N$ correspondingly. Obviously the solutions of the above equations have negative real parts.

Next we pay  attention on how perturbations  of the fields $\chi^i$ may affect the power spectrum of the curvature perturbations. 
Starting from  the action 

\bea
S_{Einstein}=\int{d^4x\sqrt{-g} \lt[\fr{R}{2}-\fr{1}{2}G_{IJ}g^{\mu\nu}\pt_{\mu}z^I \pt_{\nu}z^J-V(z^I)  \rt]   }
\eea

with respect to $z^I$ we get the equations of motion

\bea
g^{\mu\nu}z^I_{;\mu;\nu}+g^{\mu\nu}\Gamma^I_{JK}\pt_{\mu}z^J \pt_{\nu}z^K-G^{IK}V_{,K}=0 \label{eqmot}
\eea

with $\Gamma^I_{JK}(z^L)$ the Christoffel symbol in field space.
 Following  \cite{Kaiser2012} we perturb each scalar around its background value

\bea
z^I(x^\mu)=\varphi^I(t)+\delta\phi^I(x^\mu)
\eea

and correspondingly the metric around the FRW metric \cite{Mukhanov1990,Bassett2005,Malik2008}

\bea
ds^2=-(1+2A)dt^2+2a(\pt_i B)dx^i dt+a^2\lt[(1-2\psi)\delta_{ij}+2\pt_i\pt_j E\rt]dx^i dx^j.
\eea

We then consider linear perturbations of $\delta \phi^i=\mathcal{Q}^I+\mathcal{O}(\mathcal{Q}^2)$ and introducing the gauge-invariant Mukhanov-Sasaki variables \cite{Mukhanov1990,Bassett2005,Malik2008}

\bea
Q^I\equiv\mathcal{Q}^I+\fr{\dot{\varphi}^I}{H}\psi.
\eea

 eq. (\ref{eqmot}) splits into background and perturbation equations

\bea
\mathcal{D}_t \dot{\varphi}^I+3H\dot{\varphi}^I+G^{IK}V_{,K}&=&0 \label{backgr} \nonumber \\
\mathcal{D}_{t}^2 Q^I+3H\mathcal{D}_{t}Q^I+\lt[\fr{k^2}{a^2}\delta^I_J+\mathcal{M}^I_J-\fr{1}{a^3}\mathcal{D}_t \lt(\fr{a^3}{H}\dot{\varphi}^I\dot{\varphi}_J\rt)\rt]Q^J&=&0
\eea

where $\mathcal{D}_t$ denotes the covariant derivative in the field space and
\bea
\mathcal{M}^I_J\equiv G^{IK}\, (\mathcal{D}_J\mathcal{D}_KV)-\mathcal{R}^I_{LMJ}\dot{\varphi}^L\dot{\varphi}^M
\eea
with $\mathcal{R}^I_{LMJ}$ the curvature tensor in this space.

If we define $|\dot{\varphi}^I|\equiv \dot{\sigma}=\sqrt{G_{IJ}\dot{\varphi}^I\dot{\varphi}^J}$ and the unit vector along the background orbit

\bea
\hat{\sigma}^I\equiv \fr{\dot{\varphi}^I}{\dot{\sigma}}. \label{sigmadef}
\eea

and

\bea
\omega^I\equiv\mathcal{D}_t\hat{\sigma}^I ,\; \hat{s}^I\equiv \fr{\omega^I}{\omega}, \quad \omega \equiv \sqrt{\omega^I \omega_I} \label{omdef} \\
\gamma^{IJ}\equiv G^{IJ}-\hat{\sigma}^I\hat{\sigma}^J-\hat{s}^I\hat{s}^J
\eea

denoting the vectors in the perpendicular directions we get

\bea
\mathcal{D}_t \hat{s}^I=-\omega \hat{\sigma}^I-\Pi^I
\eea

and

\bea
\Pi^I\equiv \fr{1}{\omega}\mathcal{M}_{\sigma K}\gamma^{IK}, \quad \mathcal{M}_{\sigma K} \equiv   \hat{\sigma}_I\mathcal{M}^I_K.
\eea

Furhtermore the relations

\bea
\hat{\sigma}_I\Pi^I=\hat{s}_I\Pi^I=0
\eea

hold meaning $\Pi^I$ is perpendicular to the plane spanned by $\hat{\sigma}^I,\hat{s}^I.$ In our case $\Pi^I=0$ because $\hat{\sigma},\hat{s}^I$ lie on a steady plane of the coordinates $Re(T),Re(C)$ (or equivalently of $x,y$) and therefore the variation with time of $\hat{s}^I$ lies on that plane also. 

Using the above relations the equations of motion for the adiabatic (parallel)
\[
Q_{\sigma} \,=\, 
\hat{\sigma}_I \,Q^I
\]

  and the entropy (perpendicular )perturbations
\[
Q_{s} \,=\, 
\hat{s}_I \,Q^I
\]
become
\bea
\ddot{Q}_{\sigma}+3H\dot{Q}_{\sigma}+\lt[\fr{k^2}{a^2}+\mathcal{M}_{\sigma\sigma}-\omega^2-\fr{1}{a^3}\fr{d}{dt}\lt(\fr{a^3\dot{\sigma}^2}{H}\rt)\rt]Q_{\sigma}= \nonumber \\
=2\fr{d}{dt}\lt(\omega Q_s\rt)-2\lt(\fr{V_{,\sigma} }{\dot{\sigma}}+\fr{\dot{H}}{H}\rt)\lt(\omega Q_s\rt) \label{adpert}, \quad V_{,\sigma}=V_{,I}\hat{\sigma}^I
\eea

and 

\bea
\ddot{Q}_s+3H\dot{Q}_s+\lt[\fr{k^2}{a^2}+\mathcal{M}_{ss}+3\omega^2-\Pi^2 \rt]Q_s-4\fr{\omega}{\dot{\sigma}}\fr{\dot{H}}{H}\lt(\fr{d}{dt}\lt(\fr{H}{\dot{\sigma}}\dot{Q}_{\sigma}\rt)-\fr{2H}{\dot{\sigma}}\omega Q_s\rt)=\nonumber \\
=-\mathcal{D}_t\lt(\Pi_J B^J\rt)-\Pi_J\mathcal{D}_tB^J-\mathcal{M}_{sJ}B^J-3H\lt(\Pi_J B^J\rt) \label{entpert}.
\eea
correspondingly. Note that  $B_I$ denote the projection of the fields along the six directions  in which the background fields take  zero value. The reduced metric in this space is $\gamma^{IJ}$.

We note that $\fr{\dot{H}}{H}\lt(\fr{d}{dt}\lt(\fr{H}{\dot{\sigma}}\dot{Q}_{\sigma}\rt)-\fr{2H}{\dot{\sigma}}\omega Q_s\rt)=\fr{k^2}{a^2}\Psi$ where $\Psi$ is the Bardeen potential \cite{Bardeen} and this term is neglected if $k<<aH$ (we work in the super-horizon limit).
Furthermore we observe that if $\Pi^I=0$ and $\mathcal{M}_{sJ}B^J=0$, then the fields $B^I$ don't affect the evolution of $Q_s$ and $Q_{\sigma}$. This is the case in our model since  we have already discussed that $\Pi^I=0$  as the vectors $\hat{\sigma}^I,\hat{s}^I$ lie on a steady plane along the motion. 
 However the matrix $M_{IJ}$ has zero elements $M_{\sigma J},M_{s J}$ for $J$ along these directions. Therefore $M_{s J}B^J=0$ and the six $B^I$ fields don't affect the entropy perturbations which on their turn affect the adiabatic perturbations. Therefore, the power spectrum of the curvature perturbations is not affected by the perturbations   $B^I$ of the six fields $\{ImT,ReQ,ImQ,ImC,Re\Phi,Im\Phi\}$.

\section{Calculation of the spectral index and tensor to scalar ratio} \label{app2}

\subsection{The power spectra}

In this Appendix we follow closely and repeat for completeness the formulation in \cite{Byrnes2006, Lalak}, applied in the pertinent model. We note that the action of our model can be cast in the form

\bea
S=\int{d^4x\sqrt{-g}\lt[\fr{R}{2}-\fr{1}{2}(\pt_{\mu}\phi) (\pt^{\mu}\phi)-\fr{e^{2b(\phi)}}{2}(\pt_{\mu}\chi)(\pt^{\mu}\chi)-V(\phi,\chi) \rt]  }  \label{action2}
\eea

with the redefinition

\bea
x(\phi)\equiv e^{\sqrt{\fr{2}{3}}\phi}, \quad y\equiv \chi \quad \text{and}   \quad b(\phi)=\fr{\phi}{\sqrt{6}}+\fr{1}{2}ln\lt(12 \alpha c_{as}^2\rt)
\eea
 and then the background motion of the fields is given by 

\bea
\ddot{\phi}+3H\dot{\phi}+V_{\phi}&=&b_{\phi}e^{2b}\dot{\chi}^2 \nonumber \\
\ddot{\chi}+(3H+2b_{\phi} \dot{\phi})\dot{\chi}+e^{-2b}V_{\chi}&=&0
\eea

and the Friedmann equations read

\bea
\dot{H}&=&-\fr{1}{2}\lt[\dot{\phi}^2+e^{2b}\dot{\chi}^2\rt], \nonumber \\
H^2&=&\fr{1}{3}\lt[\fr{\dot{\phi}^2}{2}+\fr{e^{2b}}{2}\dot{\chi}^2+V \rt].
\eea

We will study the  perturbations in the longitudinal gauge in which the perturbed metric is given by 

\bea
ds^2=-(1+2\Phi)dt^2+a^2(1-2\Phi)dx^2
\eea

and the scalar fields $\phi,\chi$ are perturbed as

\bea
\phi(t,x)&=&\phi(t)+\delta \phi(t,x) \nonumber \\
\chi(t,x)&=&\chi(t)+\delta \chi(t,x)
\eea

The study is mainly performed in the ''rotated'' basis

\bea
\delta\sigma &\equiv& cos\theta \delta\phi+sin \theta e^b \delta\chi \nonumber \\
\delta s &\equiv& -sin\theta \delta \phi+cos \theta e^b \delta \chi
\eea 

where

\bea
cos\theta=\fr{\dot{\phi}}{\dot{\sigma}} \:, \: sin\theta=\fr{\dot{\chi}e^b}{\dot{\sigma}} \:,\: \dot{\sigma}=\sqrt{\dot{\phi}^2+e^{2b}\dot{\chi}^2},
\eea

and in the gauge invariant Mukhanov-Sasaki variables \cite{Mukhanov1990,Bassett2005,Malik2008}, defined by

\bea
Q_{\sigma}\equiv\delta\sigma-\fr{\dot{\sigma}}{H}\Phi.
\eea

In this basis the background equations become

\bea
\ddot{\sigma}+3H\dot{\sigma}+V_{\sigma}&=&0 \nonumber \\
\dot{\theta}+\fr{V_s}{\dot{\sigma}}+b_{\phi}\dot{\sigma} sin\theta&=&0,
\eea

and the perturbations can be written as

\bea
\lt( 
\begin{array}{c}
\ddot{Q_{\sigma}}  \\
\ddot{\delta s}  \end{array}
\rt)+\lt(
\begin{array}{cc}
3H &  \fr{2V_{,s}}{\dot{\sigma}} \\
-\fr{2V_{,s}}{\dot{\sigma}} & 3H \end{array}
\rt) \lt(
\begin{array}{c}
\dot{Q_{\sigma}}  \\
\dot{\delta s}  \end{array}
\rt)+\lt[
\fr{k^2}{a^2}\bf{1}+\lt( \begin{array}{cc} 
C_{\sigma\sigma} & C_{\sigma s}\\ 
C_{s \sigma} & C_{ss}  \end{array}
\rt)
\rt] \lt(
\begin{array}{c}
Q_{\sigma} \\
\delta s
\end{array}
\rt). \label{perteq}
\eea

The coefficients $C_{\sigma\sigma}, C_{\sigma s}, C_{s\sigma}, C_{ss}$ are given by

\bea
C_{\sigma\sigma}&=&V_{\sigma\sigma}-\lt(\fr{V_s}{\dot{\sigma}}\rt)^2+2\fr{\dot{\sigma}V_{\sigma}}{H}+3\dot{\sigma}^2-\fr{\dot{\sigma}^4}{2H^2}-b_{\phi}\lt(s_{\theta}^2c_{\theta}V_{\sigma}+\lt(c_{\theta}^2+1\rt)s_{\theta} V_s \rt) \nonumber \\
C_{\sigma s}&=&6H\fr{V_s}{\dot{\sigma}}+2\fr{V_{\sigma}V_s}{\dot{\sigma}^2}+2V_{\sigma s}+\fr{\dot{\sigma}V_s}{H}+2b_{\phi}\lt(s_{\theta}^3V_{\sigma}-c_{\theta}^3 V_s\rt) \nonumber \\
C_{s\sigma}&=&-6H\fr{V_s}{\dot{\sigma}}-2\fr{V_{\sigma}V_s}{\dot{\sigma}^2}+\fr{\dot{\sigma}V_s}{H} \nonumber \\
C_{ss}&=&V_{ss}-\lt(\fr{V_s}{\dot{\sigma}}\rt)^2+b_{\phi}\lt(1+s_{\theta}^2\rt)c_{\theta}V_{\sigma}+b_{\phi}c_{\theta}^2 s_{\theta}V_s-\dot{\sigma}^2 b_{\phi}^2
\eea
where we have used the fact that $b_{\phi\phi}=0$ for our model and  $s_{\theta}=sin\theta, c_{\theta}=cos\theta$  and 
\bea
V_{\sigma}&=&\hat{\sigma}^IV_I\; , \; V_{s}=\hat{s}^IV_I \nonumber \\
V_{\sigma\sigma}&=&\hat{\sigma}^I \hat{\sigma}^JV_{IJ} \; , \; V_{\sigma s}=\hat{\sigma}^I\hat{s}^JV_{IJ}\; , \; V_{ss}=\hat{s}^I\hat{s}^JV_{IJ}. \label{Vssdef}
\eea
with
\bea
\hat{\sigma}^I = (cos\theta,e^{-b}sin\theta) \; , \; \hat{s}^I=(-sin\theta,e^{-b}cos\theta)\;, \; I=\{\phi,\chi\}.
\eea
 the adiabatic and entropy ''vectors'' in field space.
Also with the substitutions $u_{\sigma}=aQ_{\sigma}$ and $u_s=a\delta s$ the equations (\ref{perteq}) read in conformal time $\tau$ , defined by $d\tau=\fr{dt}{a}$:

\bea
\lt[\lt( \fr{d^2}{d\tau^2}+k^2-\fr{a''}{a}\rt)\mathbf{1}+2\mathbf{S}\fr{d}{d\tau}+\mathbf{P}\rt]\lt(
\begin{array}{c}
u_\sigma \\
u_s
\end{array}
\rt)=0 \label{eqconf}
\eea

where 

\bea
\mathbf{S}&=&\lt(
\begin{array}{cc}
0 & \fr{aV_s}{\dot{\sigma}} \\
-\fr{aV_s}{\dot{\sigma}} & 0
\end{array}
\rt)  \nonumber \\
\mathbf{P}&=&\lt(
\begin{array}{cc}
a^2C_{\sigma\sigma} & a^2 C_{\sigma s}-2\fr{a'V_s}{\dot{\sigma}} \\
a^2 C_{s \sigma}+2\fr{a'V_s}{\dot{\sigma}} & a^2 C_{ss} 
\end{array}
\rt). \label{spmatr}
\eea

In the slow roll approximation we have

\bea
\fr{V_s}{\dot{\sigma}}=H\eta_{\sigma s}-b_{\phi}\dot{\sigma}s_{\theta}^3.
\eea
 
Furthermore, using the exact relation $a'=a^2 H$, and the quantity $\fr{a''}{a}$ up to order $x^2$ is given by

\bea
\fr{a''}{a}\approx\fr{1}{\tau^2}\lt(2+3\epsilon\rt)=\fr{1}{\tau^2}\lt(2+4(1+2 \rho c_{as}^2)^2 x^2+\mathcal{O}(x^3)\rt),
\eea

and $a$  up to order $x^2$ is given by

\bea
a\approx\fr{-\lt(1+\epsilon\rt)}{H \tau}=\fr{-\lt(1+\fr{4}{3}(1+2 \rho c_{as}^2)^2x^2+\mathcal{O}(x^3)\rt)}{H \tau}
\eea

where the slow roll parameters are given by
\be
\epsilon=-\fr{\dot{H}}{H^2}, \quad \quad
\eta_{IJ}=\fr{V_{IJ}}{3H^2}.
\ee
Then the equations (\ref{eqconf}) read 

\bea
\lt[\lt( \fr{d^2}{d\tau^2}+k^2-\fr{2+3\epsilon}{\tau^2}\rt)\mathbf{1}+2\mathbf{E}\fr{1}{\tau}\fr{d}{d\tau}+\mathbf{M}\fr{1}{\tau^2}\rt]\lt(
\begin{array}{c}
u_\sigma \\
u_s
\end{array}
\rt)=0 \label{eqconf2}
\eea

with

\bea
\mathbf{E}&=&\lt(\begin{array}{cc}
0 & -\eta_{\sigma s}+\xi s_{\theta}^3 \\
\eta_{\sigma s}-\xi s_{\theta}^3 & 0 \end{array}\rt) \nonumber \\
\mathbf{M}&=&\lt(\begin{array}{cc}
-6\epsilon+3\eta_{\sigma\sigma}+3\xi s_{\theta}^2c_{\theta} & 4\eta_{\sigma s}-4\xi s_{\theta}^3 \\
2\eta_{\sigma s}-2\xi s_{\theta}^3 & 3\eta_{ss}-3\xi c_{\theta}(1+s_{\theta}^2)
\end{array} \rt)
\eea

and $\xi=\sqrt{2}b_{\phi}\sqrt{\epsilon}$.

The above system  is of the form

\bea
u''+2\mathbf{L}u'+\mathbf{Q}u=0.
\eea

with $\mathbf{L}$ and $\mathbf{Q}$ determined appropriately from (\ref{eqconf2}).
Introducing a time-dependent orthogonal matrix $\mathbf{R}$ which satisfies $\mathbf{R}'=-\mathbf{LR}.$  and with the change of variables $u=\mathbf{R}\upsilon$ we obtain 

\bea
\upsilon''+\mathbf{R}^{-1}\lt(-\mathbf{L}^2-\mathbf{L}'+\mathbf{Q}\rt)\mathbf{R}\upsilon=0. \label{upseq}
\eea

The matrix $\mathbf{L}^2=\fr{\mathbf{E}^2}{\tau^2}$ is quadratic in the slow roll parameters in $\mathbf{E}$ so it is much smaller than $\fr{\mathbf{E}}{\tau^2}.$ Furthermore $\mathbf{L}'=-\fr{\mathbf{E}}{\tau^2}+\fr{\mathbf{E}'}{\tau}$ and since the slow roll parameters in $\mathbf{E}$ vary very slowly with time therefore we get

\bea
-\mathbf{L}^2-\mathbf{L}'\approx\fr{\mathbf{E}}{\tau^2}.
\eea

Then the part $-\mathbf{L}^2-\mathbf{L}'+\mathbf{Q}$ apart from the part proportional to the identity matrix contains

\bea
\fr{1}{\tau^2}\lt(\mathbf{E}+\mathbf{M}\rt)=\fr{3}{\tau^2}\lt(\begin{array}{cc}
-2\epsilon+\eta_{\sigma\sigma}+\xi s_{\theta}^2c_{\theta} & \eta_{\sigma s}-\xi s_{\theta}^3 \\
\eta_{\sigma s}-\xi s_{\theta}^3 & \eta_{ss}-\xi c_{\theta}(1+s_{\theta}^2)
\end{array}\rt)
\eea

which is a symmetric matrix and can be diagonalized by a matrix $\mathbf{\tilde{R}}_{*}$ at Hubble crossing 

\bea
\mathbf{\tilde{R}}_{*}=\lt(\begin{array}{cc}
cos\Theta_{*} & -sin\Theta_{*} \\
sin\Theta_{*} & cos\Theta_{*} 
\end{array}\rt)
\eea

yielding

\bea
\mathbf{\tilde{R}}_{*}^{-1}(\mathbf{M}+\mathbf{E})\mathbf{\tilde{R}}_{*}=\lt(\begin{array}{cc}
\tilde{\lambda}_1 & 0 \\
0 & \tilde{\lambda}_2 \end{array}\rt)
\eea

We note that in our model:

\bea
\xi&=& \frac{2}{3} \left(2 \rho c_{as}^2+1\right) x+\mathcal{O}(x^2)\nonumber \\
\epsilon&=&\frac{4}{3} \left(2 \rho c_{as}^2+1\right)^2 x^2+\mathcal{O}(x^3) \nonumber \\
\eta_{\sigma\sigma} &=& -\frac{4}{3} \left(2 \rho c_{as}^2+1\right)x+\mathcal{O}(x^2)\nonumber \\
\eta_{\sigma s}&=& \frac{64 \sqrt{2} \beta ^2 L  c_{as} \sqrt{f_2+\rho} \left(2 \rho c_{as}^2+1\right)}{3 (L+1)^2 \rho}x^{5/2}+\mathcal{O}(x^{7/2}) \nonumber \\
\eta_{ss}&=& \frac{2 \rho}{3 \left(f_2+\rho\right)}+\frac{ \left(2 (L+1)^2 \rho \left(2 \rho c_{as}^2+1\right)+8 \beta ^2 L\right)}{3 (L+1)^2 \left(f_2+\rho \right)}x+\mathcal{O}(x^2)\nonumber \\
c_{\theta}&=&1-\frac{64 \beta ^4 L^2  c_{as}^2 \left(f_2+\rho \right)}{(L+1)^4 \rho^2}x^3+\mathcal{O}(x^4)\nonumber \\
s_{\theta}&=& -\frac{8 \sqrt{2} \beta ^2 L  c_{as} \sqrt{f_2+\rho}}{(L+1)^2 \rho}x^{3/2}+\mathcal{O}(x^{5/2})
\eea

and therefore 

\bea
\mathbf{M}+\mathbf{E}=\lt(\begin{array}{cc} 
\mathcal{O}(x) & \mathcal{O}(x^{5/2}) \\
\mathcal{O}(x^{5/2}) & \mathcal{O}(1)
\end{array}\rt),
\eea

so $cos\Theta_{*}=1+\mathcal{O}(x^5) , sin\Theta_{*}=\mathcal{O}(x^{5/2})$.

By introducing around Hubble crossing 

\bea
w=\mathbf{\tilde{R}}_{*}^{-1}\mathbf{R}_{*}\upsilon
\eea

the system of equations (\ref{upseq}) decouple and become

\bea
w_{A}''+\lt[k^2-\fr{1}{\tau^2}(2+3\lambda_A)\rt]w_A=0, \: \text{with} \: A= 1, 2 \label{weqs}
\eea

with

\bea
\lambda_A=\epsilon-\fr{1}{3}\tilde{\lambda}_A.
\eea

The solution of (\ref{weqs}) with the appropriate asymptotic behaviour is

\bea
w_A=\fr{\sqrt{\pi}}{2}e^{i(\mu_A+\fr{1}{2})\fr{\pi}{2}}\sqrt{-\tau}H^{(1)}_{\mu_A}(-k\tau)e_A(k)
\eea

where $H_{\mu_A}^{(1)}$ is the Hankel function of the first kind of order,  $\mu_A=\sqrt{\fr{9}{4}+3\lambda_A}$ and the $e_A(k)$ are two normalised Gaussian random variables. 

By the regular definition of the power spectra 

\bea
\langle Q_A(k)Q_B(k') \rangle=8\pi^3\delta^{(3)}(k+k')\fr{2\pi^2}{k^3}P_{AB}(|k|)
\eea

and the independence of variables $w_1,w_2$ we have

\bea
a^2\langle Q_{\sigma}^\dagger Q_{\sigma}\rangle &=&cos^2\Theta_{*} \langle w_1^{\dagger}w_1\rangle + sin^2\Theta_{*}\langle w_2^{\dagger}w_2\rangle \label{qsqs}  \\
a^2\langle\delta s^{\dagger}Q_{\sigma} \rangle&=&\fr{1}{2}sin2\Theta_{*}\lt[\langle w_1^{\dagger}w_1\rangle-\langle w_2^{\dagger}w_2\rangle\rt]   \\
a^2\langle \delta s^\dagger \delta s\rangle &=&sin^2\Theta_{*} \langle w_1^{\dagger}w_1\rangle + cos^2\Theta_{*}\langle w_2^{\dagger}w_2\rangle 
\eea

where we substitute 

\bea
\langle w^{\dagger}_A w_A\rangle=\fr{\pi}{4}\lt(-\tau\rt) |H_{\mu_A}^{(1)}(-k\tau)|^2\equiv\fr{1}{2k}\fr{1}{(k\tau)^2}\mathcal{F}_A(-k\tau).
\eea

Then, by noting that 

\bea
\mathcal{R}\equiv\fr{H}{\dot{\sigma}}Q_{\sigma} \nonumber \\
\mathcal{S}\equiv \fr{H}{\dot{\sigma}}Q_{s}
\eea

we have

\bea
\mathcal{P}_{\mathcal{R}_*}&=&\lt(\fr{H_{*}^2}{2\pi\dot{\sigma}_{*}}\rt)^2(1-2\epsilon_{*})\lt[cos^2\Theta_{*}\mathcal{F}_1(-k\tau)+sin^2\Theta_{*}\mathcal{F}_2(-k\tau)\rt]  \label{pr1}\\
\mathcal{C}_{\mathcal{RS}_*}&=&\lt(\fr{H_{*}^2}{2\pi\dot{\sigma}_{*}}\rt)^2(1-2\epsilon_{*}) \fr{sin 2 \Theta_{*}}{2}\lt[\mathcal{F}_1(-k\tau)-\mathcal{F}_2(-k\tau)\rt] \label{crs} \\
\mathcal{P}_{\mathcal{S}_*}&=&\lt(\fr{H_{*}^2}{2\pi\dot{\sigma}_{*}}\rt)^2(1-2\epsilon_{*}) \lt[sin^2\Theta_{*}\mathcal{F}_1(-k\tau)+cos^2\Theta_{*}\mathcal{F}_2(-k\tau)\rt] \label{ps}
\eea

Here we mention that $cos^2\Theta=1+\mathcal{O}(x^5)$ whereas $sin^2\Theta=\mathcal{O}(x^5).$
Therefore, if we are to keep terms of order $\mathcal{O}(1),\mathcal{O}(x^1)$ in the square brackets for $\mathcal{P}_{\mathcal{R}_*}$ as leading and subleading terms then we take $cos\Theta_{*}\approx 1,\sin\Theta\approx 0$ and we have 

\bea
\mathcal{P}_{\mathcal{R}_*}=\lt(\fr{H_{*}^2}{2\pi\dot{\sigma}_{*}}\rt)^2(1-2\epsilon_{*})\mathcal{F}_1(-k\tau).
\eea

Due to the fact that $\lambda_1=\epsilon-\tilde{\lambda}_1=\mathcal{O}(x)<<1$ we can expand $\mu_1\approx\fr{3}{2}+\lambda_1$. \footnote{Here we note that $\eta_{ss}=\mathcal{O}(1)$ i.e. large and therefore we couldn' t expand $\mu_2$ around $\fr{3}{2}.$ However, it wasn' t necessary since the system up to the leading and subleading term of the curvature perturbations was diagonal and we needed only $\mu_1$ expanded around $\fr{3}{2}$ since $\lambda_1$ is small.}
The function $\mathcal{F}_1(\chi)$ can be expanded as

\bea
\mathcal{F}_1(\chi)=\fr{\pi}{2}\chi^3|H_{3/2}(\chi)|^2\lt(1+2\lambda_1 g(\chi)\rt)=(1+\chi^2)\lt(1+2\lambda_1g(\chi)\rt),
\eea

with

\bea
g(\chi)=Re\lt(\fr{1}{H_{3/2}^{(1)}(\chi)}\fr{dH_{\mu}^{(1)}(\chi)}{d\mu}\biggl|_{\mu=3/2}\rt).
\eea

Then we get

\bea
\mathcal{P}_{\mathcal{R}_*}&=&\lt(\fr{H_{*}^2}{2\pi\dot{\sigma}_{*}}\rt)^2\lt(1+k^2\tau^2\rt)\lt(1-2\epsilon_{*}\rt)\lt(1+2\lt(\epsilon_{*}-\lt(-2\epsilon+\eta_{\sigma\sigma}+\xi s_{\theta}^2c_{\theta}\rt)\rt)g\lt(\fr{k}{aH_{*}}\rt)\rt)\Rightarrow\nonumber \\
\mathcal{P}_{\mathcal{R}_*}&=&\lt(\fr{H_{*}^2}{2\pi\dot{\sigma}_{*}}\rt)^2\lt(1+k^2\tau^2\rt)\lt[1-2\epsilon_{*}+\lt(6\epsilon_{*}-\eta_{\sigma\sigma *}-2\xi_{*}s_{\theta *}c_{\theta *}\rt) g\lt(\fr{k}{aH_{*}}\rt)\rt]
\eea

Assuming  the ''constant slow roll approximation'' \cite{Lalak}, that is the slow roll parameters remain constant for few efoldings after the Hubble crossing
 we can take the limit $k\tau\rightarrow 0,\fr{k}{aH_*}\rightarrow 0$ and we have

\bea
\mathcal{P}_{\mathcal{R}_*}&=&\lt(\fr{H_{*}^2}{2\pi\dot{\sigma}_{*}}\rt)^2\lt[1-2\epsilon_{*}+\lt(6\epsilon_{*}-\eta_{\sigma\sigma *}-2\xi_{*}s_{\theta *}c_{\theta *}\rt) g\lt(0\rt)\rt] \Rightarrow\nonumber \\
\mathcal{P}_{\mathcal{R}_*}&=&\frac{(6 g(0)-7) (L+1)^2 \left(2 \rho c_{as}^2+1\right){}^2+72 \beta ^2 L c_{as}^2}{\left[64 \pi ^2 \beta  L (L+1)  \left(2 \rho c_{as}^2+1\right)^3\right]x}+\frac{9 (L+1)}{\left[256\pi^2 \beta  L  \left(2  \rho c_{as}^2+1\right){}^2\right]x^2} \label{pr2}
\eea

and since $\eta_{ss}>0,$ and it is of order $\eta_{ss}=\mathcal{O}(1)$ then in the limit $k\tau\rightarrow 0$

\bea
\mathcal{F}_{2}(-k\tau)\rightarrow 0 
\eea

and therefore we get the single field result from the adiabatic perturbations. Then, 

\bea
\mathcal{C}_{\mathcal{RS}_*}=\mathcal{O}(x^{1/2}) \nonumber \\
\mathcal{P}_{\mathcal{S}_*}=\mathcal{O}(x^3).
\eea

After Hubble crossing following \cite{Lalak}  we can write eqs (\ref{perteq})

\bea
\dot{Q}_{\sigma}&\approx& A' H Q_{\sigma}+B' H\delta s \nonumber \\
\dot{\delta s}&\approx& D' H \delta s,
\eea

with 

\bea
A'&=&-\eta_{\sigma\sigma}+2\epsilon-\xi c_{\theta}s_{\theta}^2 \nonumber \\
B'&=& -2\eta_{\sigma s}+2\xi s_{\theta}^3 \nonumber \\
D'&=&-\eta_{ss}+\xi c_{\theta}(1+s_{\theta}^2)
\eea

The integration of the above differential equations gives \cite{vandeBruck},   \cite{Wang2016,DiMarco,Avgoustidis2011} : 

\bea
Q_{\sigma}(N) &\approx& e^{\int_{N_{*}}^N{A'dN}}\lt(Q_{\sigma *}+\delta s_{*}\int_{N_{*}}^{N}{B' e^{\tilde{\gamma}}dN}\rt) \nonumber \\
\delta s(N) &\approx& \delta s_{*} e^{\int_{N_{*}}^N{D' dN}},
\eea

with $\tilde{\gamma}=\int_{N_{*}}^{N}{(D'-A') dN}.$
Then the power spectrum $\mathcal{P}_{\mathcal{R}}$ becomes 

\bea
\mathcal{P}_{\mathcal{R}}(N)\approx \mathcal{P}_{\mathcal{R}*}+\mathcal{P}_{\mathcal{S}*}\lt(\int_{N_{*}}^{N}{B' e^{\tilde{\gamma}}dN}\rt)^2+2Re\lt(\mathcal{C}_{\mathcal{RS}*}\rt)\int_{N_{*}}^{N}{B'e^{\tilde{\gamma}}dN}
\eea

Therefore we have to calculate the leading order of $\int_{N_{*}}^{N}{B' e^{\tilde{\gamma}}dN}$ and see if it changes the spectral index evaluated a few efolds after Hubble crossing $\mathcal{P}_{\mathcal{R}*}.$
Then $(D'-A')H=-\eta_{ss}H=\mathcal{O}(1)\equiv -\Gamma$ in leading order in $x$ and we get

\bea
\tilde{\gamma}=\int_{x_{*}}^{x}{(D'-A') H \fr{1}{\dot{x}}dx}=-\Gamma\int_{x_{*}}^{x}{\fr{1}{v_2 x^2}dx}=\fr{\Gamma}{v_2}\lt(\fr{1}{x}-\fr{1}{x_{*}}\rt),
\eea

where we have used $\dot{x}\propto x^2$ by (\ref{dotxdoty}).
Then $B'$ is of order $\mathcal{O}(x^{5/2})$ in leading order in $x$ and we get 

\bea
\int_{x_{*}}^{x_{end}}{B'e^{\tilde{\gamma}}dN}&\sim&\int_{x_{*}}^{x_{end}}{x^{5/2} e^{\fr{\Gamma}{v_2}(\fr{1}{x}-\fr{1}{x_{*}})}\fr{1}{\dot{x}}Hdx}\sim  \int_{x_{*}}^{x_{end}}{\fr{x^{5/2}}{v_2 x^2} e^{\fr{\Gamma}{v_2}(\fr{1}{x}-\fr{1}{x_{*}})}Hdx}\sim \nonumber \\
&\sim& \int_{x_{*}}^{x_{end}}{x^{1/2} e^{\Delta(\fr{1}{x}-\fr{1}{x_{*}})}dx}=\nonumber \\
&=&\frac{2}{3} \left(4 \Delta ^{3/2} F\left(\frac{\sqrt{\Delta }}{\sqrt{x_{*}}}\right) 
+ \left(\sqrt{x_{end}} (2 \Delta +x_{end})-4 \Delta ^{3/2} F\left(\frac{\sqrt{\Delta }}{\sqrt{x_{end}}}\right)\right)
   e^{\Delta  \left(\frac{1}{x_{end}}-\frac{1}{x_{*}}\right)}-\sqrt{x_{*}} (2 \Delta +x_{*})\right) \nonumber \\
&\sim&
(e^{-\fr{\Delta}{x_{*}}})+\mathcal{O}(x_{*}^{5/2}),	
\eea

with $F$ the Dawson Function and $\Delta\equiv\fr{\Gamma}{v_2}$.

From (\ref{crs}), (\ref{ps}) we observe that for $\mathcal{F}_2(-k\tau)\rightarrow 0$, $\mathcal{C}_{\mathcal{RS}}=\mathcal{O}(\mathcal{P}_{\mathcal{R}}x^{3/2})$ and $\mathcal{P}_{\mathcal{S}}=\mathcal{O}(\mathcal{P}_{\mathcal{R}} x^{3})$ and the super-Hubble evolution adds a $x^5$ to $\mathcal{P}_\mathcal{S}$ and a $x^{5/2}$ to $\mathcal{C}_\mathcal{RS}$. Therefore these terms are subleading to $\mathcal{P}_{\mathcal{R}*}$, so the spectral index can be safely calculated by $\mathcal{P}_{\mathcal{R}*}.$


\end{document}